\documentclass[preprint,draft,showpacs,aps]{revtex4}
\begin{document} 

\preprint{UQ Theory: October1992}

\title{On an interpolative Schr\"{o}dinger equation
and an alternative classical limit \footnote{Author's note: Archival version of an old preprint restored from obsolete electronic media. This was first submitted to {\em Phys. Rev. D} back in 1992 but rejected as being of ``insufficient interest". It describes a generalized dynamical system which contains an {\em exact embedding} of the classical Hamiltonian point mechanics alongside the entire non-relativistic quantum theory. The two are joined by a one-parameter family of deformed dynamics in a dimensionless parameter with $\hbar$ kept constant throughout. Originally, the work was presented as a toy model called $\lambda$--dynamics for the purpose of illustrating the absurdity of the Copenhagen Interpretation conception  of a ``classical domain" sitting alongside a ``quantum domain". The relevance of the work today is primarily mathematical. This preprint will  be superceded by a more contemporary study of this system in relation to the Renormalization Group and the connection between classical and quantum dynamics. This work posted under Creative Commons 3.0 - Attribution License.}}
\author{K. R. W. Jones}
\affiliation{Physics Department, University of Queensland,\\
St Lucia 4072, Brisbane, Australia.\\
(Revised version 13/10/92)}

\begin{abstract}
We introduce a simple deformed quantization prescription that
interpolates the classical and quantum sectors of Weinberg's
nonlinear quantum theory. The result is a novel classical limit
where $\hbar$ is kept fixed while a dimensionless mesoscopic
parameter, $\lambda\in[0,1]$, goes to zero. Unlike the standard
classical limit, which holds good up to a certain timescale,
ours is a precise limit incorporating true dynamical chaos, no
dispersion, an absence of macroscopic superpositions and a
complete recovery of the symplectic geometry of classical phase
space. We develop the formalism, and discover that energy
levels suffer a {\em generic perturbation\/}. Exactly, they
become $E(\lambda^{2}\hbar)$, where $\lambda = 1$ gives the
standard prediction. Exact interpolative eigenstates can be
similarly constructed. Unlike the linear case, these need no
longer be orthogonal. A formal solution for the interpolative
dynamics is given, and we exhibit the free particle as one
exactly soluble case. Dispersion is reduced, to vanish at
$\lambda = 0$. We conclude by discussing some possible 
empirical signatures, and explore the obstructions to 
a satisfactory physical interpretation. 
\end{abstract}
\maketitle

\section{Introduction}
It is generally thought that classical dynamics is a limiting
case of quantum dynamics. Certainly, the subclass of coherent
states admit a rigorous reduction of quantum dynamics to
classical dynamics as $\hbar\rightarrow 0$\cite{lim}. However,
as many authors have noted\cite{yaf}, this result does not hold
for {\em all\/} quantum states. To see this most clearly
imagine that we are in the deep semiclassical regime. I may
choose two coherent states centered about different points in
phase space. These follow the classical trajectories over some
finite time interval with an error, and dispersion, that can
be made  as  small as one pleases. Any linear combination of
these  is also a solution of the Schr\"{o}dinger evolution,
but it need not follow any classical trajectory\cite{sup}.

Ordinarily we solve this problem by prohibiting the appearance 
of such states at the classical level\cite{cat}. This can be
partially justified using measurement as a means  to remove
coherences\cite{dec}. For practical purposes the dilemma is of
no consequence\cite{con}. We are not forbidden to use the old
theory, {\em when appropriate\/}\cite{bor}. The problem is
thus one of consistency (for example, Ford et al.\cite{for}
argue that quantum suppression of dynamical chaos\cite{cha}
spells trouble for the correspondence principle). If quantum
theory is universal then why does it not give us a clean,
simple,  and {\em chaotic\/} reduction\cite{red}? 

In this paper we outline such a reduction. To do this we must
pay a heavy price and forsake the assumption of universality.
Keeping that which is good, we require a generalized theory
which contains both classical and quantum dynamics. The only
candidate we know of is Weinberg's nonlinear quantum 
theory\cite{sw1,sw2}. Elsewhere we used this to recast exact
Hamiltonian classical mechanics\cite{jon}. Here we develop a 
way to pass smoothly between both regimes. Our motivation is
curiosity; to find a nice way to do this, irrespective of what
it might mean. However, where possible we have attempted to
interpret the formalism as physical theory. This is fraught
with interpretative difficulty, but some generic empirical 
signatures can be extracted.

To set the scene, quantum theory is {\em superbly\/}
successful. In looking around to find trouble's mark, we can
think of no place but the classical regime (gravitation, the
most classical theory, remains the hardest uncraked nut). It
is at the interface between the microworld and the macroworld
that aesthetic dissaffection arises, for it is here that 
quantum stochasticity and measurement prove necessary. 
Most ``resolutions'', ``new interpretations'',
whatever\ldots, depart little from the orthodox theory. 
Here our philosophy is to first enlarge quantum dynamics and
then seek a natural way to blend the classical and quantum
components together. The interpolative dynamics is then put
forward as a candidate to describe a regime that borders the
cut we customarily make in everyday calculations. We make a
guess at some kind of general theoretical structure in which
to think around the questions. Without evidence that quantum
theory fails we can do no more. Why do it then? Because when
no alternative exists we are unlikely to find any failure.

\section{Classical mechanics in Weinberg's theory}
Unlike regular classical mechanics, the carbon copy within
Weinberg's  theory employs wavefunctions, $\hbar$ and the
commutation relation $[\hat{q},\hat{p}]=i\hbar$. To form
it we take any classical function, say $H(q,p)$, and turn 
it into a Weinberg observable\cite{sw3} via the ansatz 
\begin{equation}  
\label{e1} h_{0}(\psi,\psi^{*}) \equiv 
\langle\psi|H(\langle\hat{q}\rangle,
\langle\hat{p}\rangle)|\psi\rangle, 
\end{equation}
where 
$\langle\hat{q}\rangle
\equiv\langle\psi|\hat{q}|\psi\rangle/n$,
$\langle\hat{p}\rangle
\equiv\langle\psi|\hat{p}|\psi\rangle/n$, with
$n=\langle\psi|\psi\rangle$. Commutators are then
replaced by the Weinberg bracket, 
\begin{equation}
\label{e2}
[g,h]_{\rm W}
\equiv g\star h - h\star g,
\end{equation}
where $g\star h=\delta_{\psi}g\delta_{\psi^{*}}h$ and
$\delta_{\psi}$, and $\delta_{\psi^{*}}$ are shorthand for
functional derivatives\cite{fun}. Canonical commutators 
then translate to: $[\langle\hat{q}\rangle,
\langle\hat{q}\rangle]_{\rm W}=0$,
$[\langle\hat{p}\rangle,\langle\hat{p}\rangle]_{\rm W}=0$, and
$[\langle\hat{q}\rangle,\langle\hat{p}\rangle]_{\rm W}
=i\hbar/n$. The equation of motion now reads,
\begin{equation} 
\label{e3} 
i\hbar\frac{dg}{dt} = [g,h]_{\rm W}.
\end{equation}
Taking the special functionals (\ref{e1}) one
shows\cite{jon} that $[g,h]_{\rm W} = i\hbar n\{G,H\}_{\rm
PB}$. Then, since the dynamics is norm preserving, we have
$dn/dt=0$ and (\ref{e3}) reduces to 
\begin{equation}  
\label{e4}
\frac{dG}{dt}= \{G,H\}_{\rm PB}
\equiv 
\partial_{\langle\hat{q}\rangle}G
\partial_{\langle\hat{p}\rangle}H-
\partial_{\langle\hat{p}\rangle}G
\partial_{\langle\hat{q}\rangle}H.
\end{equation}
Hitherto, noncommutativity was thought to embody the essential
difference between the classical and quantum theories. Now we
see things differently, (\ref{e3}) reduces to (\ref{e4}) for
any value of $\hbar$.

What, then, is the fundamental difference? To see this, we
simply compare the classical functional ansatz (\ref{e1}) to
the Weinberg analogue of canonical quantization, 
\begin{equation} 
\label{e5}
h_{1}(\psi,\psi^{*}) \equiv 
\langle\psi|\hat{H}(\hat{q},\hat{p})|\psi\rangle.
\end{equation}
Now equation (\ref{e3}) reduces to the familiar result
\begin{equation}
\label{e6}
i\hbar \frac{d}{dt}
\langle\psi|\hat{G}|\psi\rangle
= \langle\psi|[\hat{G},\hat{H}]|\psi\rangle.
\end{equation}
Clearly, Weinberg's theory is general enough to embrace
both standard quantum theory and a novel wave version of
Hamiltonian classical mechanics. 

Our point of departue for an alternative classical limit
is the recovery of a familiar result. Comparing (\ref{e1})
and (\ref{e5}), we write
\begin{equation}
h_{1} = h_{0}\left(1 + \frac{h_{1}-h_{0}}{h_{0}}\right).
\end{equation}
At any $\hbar$ the classical approximation is good
for those $\psi$ such that $(h_{1}-h_{0})/h_{0}\ll 1$. Two
features deserve explicit note: the smaller is $\hbar$ the
better is the approximation for a given $\psi$; and, for all
non--zero $\hbar$, there exist states such that the 
criterion fails. 

\section{An interpolative domain?}
Some functionals $h(\psi,\psi^{*})$ are classical, of form
(\ref{e1}), others are quantal, of form (\ref{e5}), while
most are neither. Since both sectors are disjoint for all
$\hbar$ we seek an interpolation which joins them. In physical
terms, we imagine that the correspondence principle is to be
taken literally. Thus we speculate that, some objects,
composed of many quantum particles, act as a collective {\em
mesoparticle\/}\cite{mes}, with a center of mass dynamics that
is neither strictly quantum nor strictly classical, but some
curious blend of both. For simplicity, we assume that a
one--particle equation can do this many--particle job. 

\section{Deformed quantization}
\subsection{The mathematical notion}
To formulate this concept we generalize the central idea of 
canonical quantization and postulate a map which sends any
classical phase space function $H(q,p)$ into a one--parameter
family of {\em interpolative\/} Weinberg observables
$h_{\lambda}(\psi,\psi^{*})$. Symbolically, we write    
\begin{equation}   
\label{e7}   
{\cal Q}^{\lambda}_{\psi}
\vdash H(q,p)\stackrel{\lambda}{\mapsto}
h_{\lambda}(\psi,\psi^{*}), 
\end{equation} 
and call ${\cal Q}^{\lambda}_{\psi}$ a {\em deformed
quantization\/}. Imposing (\ref{e1}) and (\ref{e5}) as known
boundary conditions, we interpret $\lambda\in[0,1]$ as a
dimensionless index of mesoscopic effects.

Since $\lambda$ is to govern emergence of classical behaviour
we expect it to depend upon some function of particle size,
mass, number, or mixture thereof. There is no way to guess
this. Some authors suggest that gravity could have something
to do with it\cite{gra}. Here we pick $\lambda(m)\equiv
1/(1+(m/m_{P})^{\alpha})$, for some $\alpha>0$, where
$m_{P}=2.177\times10^{-5} {\rm g}$ is the Planck mass to
illustrate how the proposal might work\cite{pla}. However, 
we emphasize that $\lambda$ is an adjustable parameter which
cannot be fixed within this framework.

\subsection{The specific proposal}
With only the boundary conditions known we cannot fix
(\ref{e7}) uniquely. However, since the ansatz (\ref{e1})
contains only expectations, and (\ref{e5}) only operators, 
it is suggestive to deform the particle coordinates via the 
simple convex combination\cite{lie}:  
\begin{eqnarray}   
\label{e8}
\hat{q}_{\lambda}& \equiv &
\lambda\hat{q} + (1-\lambda)\langle\hat{q}\rangle,\\
\label{e9}
\hat{p}_{\lambda}& \equiv &
\lambda\hat{p} +
(1-\lambda)\langle\hat{p}\rangle.
\end{eqnarray}
This prescription is unique among linear combinations 
once we impose the physical constraints:
$q_{\lambda}\equiv\langle\hat{q}_{\lambda}\rangle
=\langle\hat{q}\rangle$, and
$p_{\lambda}\equiv\langle\hat{p}_{\lambda}\rangle
=\langle\hat{p}\rangle$. These enforce invariance of both the
center of mass coordinates, and the canonical Weinberg 
bracket relations under deformation.

Having chosen the deformed operators we now select the
obvious generalization of canonical quantization:
\begin{equation} 
\label{e10}
{\cal Q}^{\lambda}_{\psi}
\vdash H(q,p)\stackrel{\lambda}{\mapsto}
h_{\lambda}(\psi,\psi^{*})
= \langle\psi|
\hat{H}^{\lambda}
|\psi\rangle,
\end{equation}
where $\hat{H}^{\lambda}\equiv
\hat{H}(\hat{q}_{\lambda},\hat{p}_{\lambda})$,
and, for definiteness, we assume that $\hat{q}_{\lambda}$
and $\hat{p}_{\lambda}$ are Weyl--ordered\cite{wey}. As we 
now show, (\ref{e10}) gives an interpolative dynamical system
with some interesting properties. For inessential simplicity
we treat only systems with one classical degree of freedom.
The generalization is straightforward.

\section{The reduced Weinberg bracket}
\subsection{A general reduction lemma}
Of fundamental importance is the effect of the ansatz
(\ref{e10}) upon the bracket (\ref{e2}). We begin with a
computation for the more general class of functionals 
\begin{equation}
\label{e11}
h(\psi,\psi^{*})\equiv \langle\psi|
\hat{H'}(\hat{q},\langle\hat{q}\rangle;
\hat{p},\langle\hat{p}\rangle)
|\psi\rangle,
\end{equation}
with $H'(q_{1},q_{2};p_{1},p_{2})$ an auxilliary $c$--number
function. Applying the chain rule first, the
functional derivative of this expands to 
\begin{eqnarray}
\lefteqn{\delta_{\psi}h
 = \langle\psi|\hat{H'} +} \nonumber \\
& &\hspace{0.7cm} \langle\psi|
\partial_{\langle\hat{q}\rangle}\hat{H'}
|\psi\rangle
\delta_{\psi}\langle\hat{q}\rangle
+\langle\psi|
\partial_{\langle\hat{p}\rangle}\hat{H'}
|\psi\rangle
\delta_{\psi}\langle\hat{p}\rangle.
\label{e12}
\end{eqnarray}
Evaluating $\delta_{\psi}\langle\hat{q}\rangle$ and
$\delta_{\psi}\langle\hat{p}\rangle$ gives the bra--like pair:
\begin{eqnarray}
\label{e13}
\delta_{\psi}\langle\hat{q}\rangle
& = & \langle\psi|(\hat{q}-\langle\hat{q}\rangle)/n,\\
\label{e14}
\delta_{\psi}\langle\hat{p}\rangle
& = & \langle\psi|(\hat{p}-\langle\hat{p}\rangle)/n.
\end{eqnarray}
Taking hermitian adjoints of 
(\ref{e12}), (\ref{e13}) and (\ref{e14})
gives the ket--like quantities $\delta_{\psi^{*}}h$, 
$\delta_{\psi^{*}}\langle\hat{q}\rangle$, and
$\delta_{\psi^{*}}\langle\hat{p}\rangle$. Using these
rules it becomes a simple matter to expand 
$$[g,h]_{\rm W}=\delta_{\psi}g\delta_{\psi^{*}}h
- \delta_{\psi^{*}}g\delta_{\psi}h.$$
In reducing the expansion it is helpful to identify
like terms and to make frequent use of (\ref{e13}),
(\ref{e14}) and their adjoints. Of special utility is 
a family of results like
$$\delta_{\psi}\langle\hat{q}\rangle\hat{H'}|\psi\rangle -
\langle\psi|\hat{H'}\delta_{\psi^{*}}\langle\hat{q}\rangle
= i\hbar\langle\psi|\partial_{\;\hat{p}\;}\hat{H}'
|\psi\rangle/n,$$
where $\partial_{\hat{q}} \equiv[\bullet,\hat{p}]/i\hbar$,
and $\partial_{\hat{p}} \equiv[\hat{q},\bullet]/i\hbar$.
Then, after some cancellation using canonical bracket
relations, and some rearrangement, we find that
\begin{eqnarray} 
\lefteqn{
[g,h]_{\rm W}
 = \langle\psi|[\hat{G'},\hat{H'}]|\psi\rangle }
\nonumber  \\ 
& & \mbox{}+ 
i\hbar\left\{
\langle\psi|\partial_{\langle\hat{q}\rangle}\hat{G'}
|\psi\rangle
\langle\psi|
\;\partial_{\hat{p}}\;\hat{H'}
|\psi\rangle -
\langle\psi|
\;\partial_{\hat{p}}\;\hat{G'}
|\psi\rangle
\langle\psi|
\partial_{\langle\hat{q}\rangle}\hat{H'}
|\psi\rangle
\right\}/n
\nonumber \\
& & \mbox{}+ 
i\hbar\left\{
\langle\psi|
\;\partial_{\hat{q}}\;\hat{G'}
|\psi\rangle
\langle\psi|
\partial_{\langle\hat{p}\rangle}\hat{H'}
|\psi\rangle -
\langle\psi|
\partial_{\langle\hat{p}\rangle}\hat{G'}
|\psi\rangle
\langle\psi|
\;\partial_{\hat{q}}\;\hat{H'}
|\psi\rangle
\right\}/n
\nonumber  \\ 
& & \mbox{}+ 
i\hbar\left\{
\langle\psi|
\partial_{\langle\hat{q}\rangle}\hat{G'}
|\psi\rangle
\langle\psi|
\partial_{\langle\hat{p}\rangle}\hat{H'}
|\psi\rangle -
\langle\psi|
\partial_{\langle\hat{p}\rangle}\hat{G'}
|\psi\rangle
\langle\psi|
\partial_{\langle\hat{q}\rangle}\hat{H'}
|\psi\rangle
\right\}/n.
\label{e15}
\end{eqnarray}
This expression is rather more general than is required, but
displays the essential origin of our next result.

\subsection{Reduction for interpolative observables}
To treat the interpolative case (\ref{e10}) we choose
$$H'=H(\lambda q_{1}+(1-\lambda)q_{2}, 
\lambda p_{1}+(1-\lambda)p_{2}).$$
Then, $\partial_{\hat{q}}\hat{H'} =
\lambda\hat{H}^{\lambda}_{q}$,
$\partial_{\langle\hat{q}\rangle}\hat{H'} =
(1-\lambda)\hat{H}^{\lambda}_{q}$,
$\partial_{\hat{p}}\hat{H'} = \lambda
\hat{H}^{\lambda}_{p}$, and
$\partial_{\langle\hat{p}\rangle}\hat{H'} =
(1-\lambda)\hat{H}^{\lambda}_{p}$,  where
$\hat{H}^{\lambda}_{q}$ and $\hat{H}^{\lambda}_{p}$ denote
the quantized classical partials of $H(q,p)$. Thus
(\ref{e15}) becomes 
\begin{eqnarray}
\lefteqn{[g_{\lambda},h_{\lambda}]_{\rm W} = 
\langle\psi|
[\hat{G}^{\lambda}
,\hat{H}^{\lambda}]
|\psi\rangle} \nonumber \\
& & \mbox{}+
i\hbar(1-\lambda^{2}) 
\left\{\langle\psi|
\hat{G}_{q}^{\lambda}
|\psi\rangle
\langle\psi|
\hat{H}_{p}^{\lambda}
|\psi\rangle 
-
\langle\psi|
\hat{G}_{p}^{\lambda}
|\psi\rangle
\langle\psi|
\hat{H}_{q}^{\lambda}
|\psi\rangle\right\}/n.
\label{e16}
\end{eqnarray}
Thus the ansatz (\ref{e10}) collects the three residual terms
of (\ref{e15}) into a ``mean--field'' Poisson
bracket\cite{mft}. The scale factor $(1-\lambda^{2})$ now
controls the mixture of quantum and classical
effects\cite{moy}.

\section{An Interpolative Schr\"{o}dinger equation}
\subsection{The equation of motion for expectation values}
Substituting (\ref{e16}) into (\ref{e3}), and using
the property that $dn/dt=0$, now gives
\begin{eqnarray}  
\lefteqn{\frac{d\langle G^{\lambda}\rangle}{dt}
\equiv \langle[\hat{G}^{\lambda}
,\hat{H}^{\lambda}]\rangle/i\hbar +} \nonumber \\
& &\hspace{1cm}(1-\lambda^{2})
\left\{\langle\hat{G}_{q}^{\lambda}\rangle
       \langle\hat{H}_{p}^{\lambda}\rangle -
       \langle\hat{G}_{p}^{\lambda}\rangle
       \langle\hat{H}_{q}^{\lambda}\rangle\right\},
\label{e17}
\end{eqnarray}
where $\langle\bullet\rangle\equiv
\langle\psi|\bullet|\psi\rangle/n$. This provides
an interpolative analogue of the standard Schr\"{o}dinger
picture equation of motion for expectation values. 

Of course, at $\lambda=0$ all deformed operators commute and
the first term vanishes. We are thus left with the second term
alone and (\ref{e4}) drops out directly. The other limit
$\lambda=1$ kills the second term, operators revert to their
standard canonical quantizations and (\ref{e6}) results. So
the commutator term is certainly ``quantum'' and the bracket
term is certainly ``classical''.

\subsection{An interpolative Ehrenfest theorem}
Applying (\ref{e17}) to the coordinate operators now gives 
an interpolative Ehrenfest--type theorem\cite{ehr}:
\begin{eqnarray}
\label{e18}
\frac{d\langle\hat{q}_{\lambda}\rangle}{dt}
& = & + \langle \hat{H}^{\lambda}_{p} \rangle \\
\label{e19}
\frac{d\langle\hat{p}_{\lambda}\rangle}{dt}
& = & - \langle \hat{H}^{\lambda}_{q} \rangle.
\end{eqnarray}
Using this we obtain valuable insight about how wave
propagation is affected by $\lambda$. For instance,
choosing $H(q,p)=p^{2}/2m + V(q)$, we find that the 
only change appears in the force term. After some
rearrangement, this reads
\begin{eqnarray}
\frac{d\langle\hat{p}_{\lambda}\rangle}{dt} & = &
- \langle V_{q}(\langle\hat{q}\rangle +
\lambda[\hat{q}-\langle\hat{q}\rangle])\rangle \nonumber \\
& = & - \sum_{k=0}^{\infty} \frac{\lambda^{k}}{k!}
\langle[\hat{q}-\langle\hat{q}\rangle]^{k}\rangle
\partial_{q}^{k+1}V(\langle\hat{q}\rangle).
\label{force}
\end{eqnarray}
Looking at this we see that $\lambda$ controls the range
at which the wavefunction $\psi$ probes the potential $V(q)$.
At the classical extreme, wavepackets feel only the classical
force at their centre, whereas, in the quantum extreme, this
is averaged over space\cite{mes}.  

\subsection{The interpolative wave equation}
Consider now Weinberg's generalized Schr\"{o}dinger 
equation\cite{sw4} 
$$i\hbar \frac{d\psi}{dt}
=\delta_{\psi^{*}} h.$$ Although nonlinear, standard Hilbert
space methods are easily adapted by using the definition
(\ref{e10}), along with the hermitian adjoints of (\ref{e12}),
(\ref{e13}) and (\ref{e14}), to introduce an {\em effective
Hamiltonian operator\/},   
\begin{eqnarray}
\lefteqn{\hat{H}_{\rm eff}^{\lambda}(\psi,\psi^{*})
\equiv \hat{H}^{\lambda} + } \nonumber\\
& & 
(1-\lambda)\left\{\langle H^{\lambda}_{q}\rangle
(\hat{q}-\langle\hat{q}\rangle) +
\langle H^{\lambda}_{p}\rangle 
(\hat{p}-\langle\hat{p}\rangle)\right\},
\label{e21}
\end{eqnarray}
such that $\delta_{\psi^{*}}h= \hat{H}_{\rm eff}^{\lambda}
(\psi,\psi^{*})|\psi\rangle$. This operator defines the 
interpolative Schr\"{o}dinger equation\cite{kib}, 
\begin{equation}
\label{e22}
i\hbar\frac{d}{dt}|\psi\rangle
= \hat{H}_{\rm eff}^{\lambda}(\psi,\psi^{*})
|\psi\rangle.
\end{equation}
One verifies easily that $\lambda =1$ returns the ordinary
linear Schr\"{o}dinger equation. However, for $\lambda\ne 1$ 
the operator is generally $\psi$--dependent. 

This property is responsible for the failure of many standard
results, such as the superposition principle, preservation of
the {\em global\/} inner product between distant states, and
the orthogonality of eigenvectors for self--adjoint operators.
Proofs of these assume that $\hat{H}$ is the {\em same\/} for
any quantum state.

Choosing $H(q,p)=p^{2}/2m + V(q)$, we set $\hat{q} = q$ and
$\hat{p} = -i\hbar\partial_{q}$. Then defining,
\begin{eqnarray}
\label{qpar}
Q(t) & = & n^{-1}\int_{-\infty}^{\infty}
\psi(q,t)^{*} q \psi(q,t)\,dq\\
\label{ppar}
P(t) & = & n^{-1}\int_{-\infty}^{\infty}
\psi(q,t)^{*} \{-i\hbar\partial_{q} \psi(q,t)\}\,dq,\\
\label{fpar}
F(t) & = & n^{-1}\int_{-\infty}^{\infty}
\psi(q,t)^{*}V_{q}(\lambda q + (1-\lambda)Q(t))
\psi(q,t)\,dq,
\end{eqnarray}
equations (\ref{e21}) and (\ref{e22}) yield the explicit
nonlinear integrodifferential wave equation,
\begin{eqnarray*}
\lefteqn{i\hbar\frac{\partial\psi(q,t)}{\partial t} = 
\frac{1}{2m} \left\{-\lambda^{2}\hbar^{2} \partial_{q}^{2}
- 2i\hbar (1- \lambda^{2}) P(t) \partial_{q}
- (1-\lambda^{2}) P^{2} (t) \right\} \psi(q,t)} \\ 
& & + \left\{V(\lambda q + (1-\lambda)Q(t)) 
+ (1-\lambda) F(t)(q - Q(t))\right\}\psi(q,t). 
\end{eqnarray*}
The nonlinearity of (\ref{e22}) lies in those terms carrying 
state dependent parameters (\ref{qpar}), (\ref{ppar}) and
(\ref{fpar}). Given the complexity of this form, abstract
operator techniques are preferable. Calculations with the
explicit equation are hideous.

Of particular interest is the case $\lambda = 0$. From
(\ref{e21}) we compute the effective classical Hamiltonian  
\begin{eqnarray}
\lefteqn{\hat{H}_{\rm eff}^{0}(\psi,\psi^{*}) \equiv 
H(\langle\hat{q}\rangle,\langle\hat{p}\rangle) +}
\nonumber \\
& & H_{q}(\langle\hat{q}\rangle,\langle\hat{p}\rangle)
(\hat{q}-\langle\hat{q}\rangle)
+H_{p}(\langle\hat{q}\rangle,\langle\hat{p}\rangle)
(\hat{p}-\langle\hat{p}\rangle).
\label{e23}
\end{eqnarray}
Combining (\ref{e22}) and (\ref{e23}) now gives us a 
classical Schr\"{o}dinger equation.

\section{A classical Schr\"{o}dinger equation}
From (\ref{e4}), we know that all solutions $\psi(t)$ must 
have expectations, $Q(t)\equiv\langle\hat{q}\rangle$ and
$P(t)\equiv\langle\hat{p}\rangle$, that precisely follow   
the classical trajectories of any chosen $H(q,p)$, for     
all time, and for all values of $\hbar$. We now construct  
the explicit solution.

\subsection{An heuristic overview}
For short time intervals, $\Delta t$, we can assume that
(\ref{e23}) is constant. In the simplest approximation,  
we let $\psi_{t_{0}}$  be the initial wavefunction, and
construct the infinitesimal {\em unitary\/} propagator
\begin{equation}   
\label{e24} 
\hat{U}_{\Delta t}\approx
\exp\left\{-\frac{i\Delta t}{\hbar}
\hat{H}_{\rm eff}^{0}
(\psi_{t_{0}},\psi^{*}_{t_{0}})\right\}.
\end{equation}
Then, since (\ref{e23}) is linear in $\hat{q}$ and
$\hat{p}$, it follows that (\ref{e24}) is a member
of the Heisenberg--Weyl group\cite{hwg}. Operators of
this type assume the general form,
\begin{equation}
\label{e25}
\hat{U}[Q,P;S]
\equiv \exp\left\{\frac{i}{\hbar}\left[S\hat{1} +
P\hat{q} - Q\hat{p}\right]\right\},
\end{equation}
and obey the operator relations:
\begin{eqnarray}
\label{e26}
\hat{U}^{\dagger}[Q,P;S]\hat{q}\hat{U}[Q,P;S]
& = & \hat{q} + Q\hat{1}, \\
\label{e27}
\hat{U}^{\dagger}[Q,P;S]\hat{p}\hat{U}[Q,P;S]
& = & \hat{p} + P\hat{1}.
\end{eqnarray}
Comparing (\ref{e24}) and (\ref{e25}), and rewriting
the definition (\ref{e23}) in the form,
\begin{equation}
\label{e28}
\hat{H}_{\rm eff}^{0}
= -\left\{Q H_{q} + P H_{p}- H\right\}\hat{1} 
+ H_{q}\hat{q} + H_{p}\hat{p},
\end{equation}
now gives the approximate result
\begin{eqnarray*}
|\psi_{t_{0}+\Delta t}\rangle
&\approx & \hat{U}_{\Delta t}|\psi_{t_{0}}\rangle\\
& = & \hat{U}[+H_{p}\Delta t,-H_{q}\Delta t;\Delta S]
|\psi_{t_{0}}\rangle,
\end{eqnarray*}
with $\Delta S =\{Q(t_{0})H_{q} + P(t_{0})H_{p} - H\}
\Delta t$. Invoking (\ref{e26}) and (\ref{e27}), it 
follows that: 
\begin{eqnarray*}
Q(t_{0}+\Delta t) & \approx & Q(t_{0}) 
+ H_{p}(Q(t_{0}),P(t_{0}))\Delta t,\\
P(t_{0}+\Delta t) & \approx & P(t_{0}) 
- H_{q}(Q(t_{0}),P(t_{0}))\Delta t.
\end{eqnarray*}
These considerations show how the effective Hamiltonian
(\ref{e23}) propagates {\em any\/} wave $\psi$ along classical
trajectories, as expected from equation (\ref{e4}).

\subsection{The exact treatment}
Suppose we construct the operator $U[Q(t),P(t);S(t)]$ using
parameters $Q(t)$ and $P(t)$ that are obtained from solving
Hamilton's equations for the initial conditions, $Q(t_{0})$,
and $P(t_{0})$. Specifically, we demand that,  
\begin{eqnarray}
\label{e29} 
\dot{P}(t) & = & -\partial_{q}H(Q,P),\\
\label{e30}
\dot{Q}(t) & = & +\partial_{p}H(Q,P),
\end{eqnarray}
for all $t\ge t_{0}$. Then, choosing $\psi_{0}$ to
be an arbitrary state with both coordinate expectation 
values equal to zero, we construct the trial solution
$$|\psi_{t}\rangle=
U[Q(t),
P(t);S(t)]|\psi_{0}\rangle,\;\;t\ge t_{0}.$$ 
Equation (\ref{e4}) is now trivially satisified. To verify
(\ref{e22}) note that $\hat{U}[t]$ determines,
\begin{equation}
\label{e31}
\hat{H}(t) \equiv i\hbar\left\{\frac{d}{dt}\hat{U}[t]\right\}
\hat{U}^{\dagger}[t].
\end{equation}
Then, using the Weyl multiplication rule\cite{hwg}, 
\begin{eqnarray}
\label{e32}
\lefteqn{\hat{U}[Q_{2},P_{2};S_{2}]
\hat{U}^{\dagger}
[Q_{1},P_{1};S_{1}] = 
e^{i/2\hbar\{P_{1}Q_{2}-Q_{1}P_{2}\}}
\times } \nonumber \\
&&\hspace{1cm}\hat{U}[Q_{2}-Q_{1},P_{2}-P_{1};S_{2}-S_{1}],
\end{eqnarray}
and (\ref{e31}), we compute:
\begin{eqnarray*}
\hat{H}_{\rm eff}^{0}
& = & i\hbar \lim_{\delta t\rightarrow 0}
\frac{\hat{U}[Q(t+\delta t),P(t+\delta t);S(t+\delta t)]
\hat{U}^{\dagger}[Q(t),P(t);S(t)]-\hat{1}}{\delta t}\\
& = & i\hbar \lim_{\delta t\rightarrow 0}
\frac{e^{i\delta t/2\hbar\{P\dot{Q}-Q\dot{P}\}}
\hat{U}^{\dagger}[\dot{Q}\delta t,\dot{P}\delta t;
\dot{S}\delta t]-\hat{1}}{\delta t} \\
& = & - \{(P\dot{Q}-Q\dot{P})/2+\dot{S}\}
 - \dot{P}\hat{q}+\dot{Q}\hat{p}.
\end{eqnarray*}
Comparing this to (\ref{e28}), we first pick out
(\ref{e29}) and (\ref{e30}) as necessary conditions. Then,
looking at the constant term, we solve for $\dot{S}$ to 
obtain $\dot{S}=1/2(P\dot{Q} - Q\dot{P}) - H$. Integrating
$\dot{S}$ now gives the exact classical propagator,
\begin{equation}
\label{e33}
\hat{U}[t] = \exp\left\{\frac{i}{\hbar}\left[
\phi(t)\hat{1} + P(t)\hat{q} -Q(t)\hat{p}\right]\right\}, 
\end{equation}
where $Q(t)$ and $P(t)$ obey (\ref{e29}), and the phase
factor $\phi(t)$ reads 
\begin{equation}
\label{e34}
\phi(t) = \int_{t_{0}}^{t}
\left(
\frac{P\dot{Q} - Q\dot{P}}{2}
\right) 
- H(Q,P)\,d\tau.
\end{equation}
Unlike ordinary classical mechanics, our wave version has an
extra degree of freedom; a phase factor. As one might have
expected\cite{act}, this phase records the classical action.
However, unlike linear theory, the phase--to--action
correspondence is now exact.

\subsection{Phase anholonomy effects}
Interestingly, (\ref{e34}) contains a simple Berry
phase\cite{ber}. To isolate this we employ the 
Aharanov--Anandan\cite{aap} formula, $\dot{\gamma(t)}=
i\langle\tilde{\psi}|\{d/dt|\tilde{\psi}\rangle\}$,
where $\tilde{\psi}$ is a ray--space trajectory. If 
$|\tilde{\psi}(0)\rangle$ is any state with vanishing
coordinate expectations, then a ray path can be
parametrized as $|\tilde{\psi}(t)\rangle = 
\hat{U}[Q(t),P(t);0]|\tilde{\psi}(0)\rangle,$
to give,
\begin{eqnarray*} 
\dot{\gamma(t)} & = &
i\langle\tilde{\psi}(0)|
\hat{U}^{\dagger}[t]
\left\{\frac{d}{dt}\hat{U}[t]\right\}
|\tilde{\psi}(0)\rangle \\
& = &\langle\tilde{\psi}(0)|
(P\dot{Q}-Q\dot{P})/2 - \dot{P}\hat{q} + \dot{Q}\hat{p}
|\tilde{\psi}(0)\rangle/\hbar \\
& = & (P\dot{Q}-Q\dot{P})/2\hbar.
\end{eqnarray*}
On a closed loop $\Gamma$, we find  
$\int_{0}^{T}
P\dot{Q}\,dt = +\oint_{\Gamma} P\,dQ$, and
$\int_{0}^{T} Q\dot{P}\,dt = -\oint_{\Gamma} P\,dQ$, where
$T$ is the circuit time and signs are fixed by the sense
of traversal. Thus, 
\begin{equation}
\label{e35}
\gamma(\Gamma) = +\frac{1}{\hbar}\oint_{\Gamma} P\,dQ.  
\end{equation}
This explicit relationship suggests that geometric phases
upon closed loops might well be interpreted as the natural
action variables of quantum mechanics.

\subsection{Explicit wavefunction solutions}
Returning to (\ref{e33}), we now seek explicit
wavefunction solutions. Passing to the Schr\"{o}dinger
representation, $\hat{q}\mapsto q$, and $\hat{p}\mapsto 
-i\hbar\partial_{q}$, we note the standard result\cite{hwg},
\begin{equation}
\label{dis}
U[Q,P;0]\psi(q) =  e^{-iPQ/2\hbar}
e^{iPq/\hbar}\psi(q - Q).
\end{equation}
Then, given any state $\psi_{0}(q)$ with both expectation
values equal to zero, equation (\ref{e33}) yields
$$\psi(q,t) =
e^{i\phi(t)/\hbar}
e^{-iP(t)Q(t)/2\hbar}
e^{iP(t)q/\hbar}\psi_{0}(q - Q(t)).$$
Looking at this we see directly that {\em all\/} waves
propagate without dispersion. The arbitrary wave envelope 
$\psi_{0}(q)$ preserves its shape while being moved around
in Hilbert space via its expectation value parameters 
$Q(t)$ and $P(t)$. Therefore, no interference or 
tunnelling is possible in this limit. A wave--packet must
reflect or pass a barrier with certainty, just as a point
particle does in ordinary classical mechanics. Suppose we 
fire a packet at a double slit. Then it must go through either
one or the other slit, or it must strike the slit screen and
return. Hence it is possible to view interference and
diffraction phenomena as products of {\em linear dynamics\/}.
Pick the right kind of nonlinear propagation, and they
evaporate altogether\cite{eva}. 

\subsection{The recovery of classical phase space}
Since the wave aspects are frozen out, we can now build a
faithful analogue of classical phase space. To define
this, we introduce the coordinate map,
$$\Pi\vdash{\cal
H}\mapsto {\bf R}^{2} \mbox{    where    } \Pi[\psi] =
(\langle\hat{q}\rangle,\langle\hat{p}\rangle).$$
The appropriate mathematical object involves a partition
of Hilbert space into disjoint sets of wavefunctions     
which share identical coordinate expectations. These sets
are defined as the $\Pi$--induced equivalence classes,
$$\tilde{\psi}(Q,P) =\{\psi\in{\cal H}\vdash
\Pi[\psi]=(Q,P)\in {\bf R}^{2}\}.$$
One can now treat the labels $(Q,P)$ as points, just like
in ordinary classical phase space. Each emblazons a bag of
$\Pi$--equivalent wavefunctions. We think of the classical
limit as a dynamical regime where $\psi$ does not matter, 
only its parameters $(Q,P)$. The original classical
Hamiltonian $H(q,p)$ now determines, via the ansatz
(\ref{e1}), and equations, (\ref{e22}), and (\ref{e23}), 
a symplectomorphism of this phase space\cite{sym,der}. 

\section{The interpolative propagator}
\label{liou}
\subsection{The Liouville equation}
Introducing a Liouville operator ${\cal L}_{h}\equiv
[\bullet,h]_{\rm W}$, such that ${\cal L}_{h}\circ g
\equiv [g,h]_{\rm W}$ with iterated ``powers'': ${\cal
L}_{h}^{k+1} \circ g = [{\cal L}_{h}^{k}\circ g ,h]_{\rm W}$,
we can obtain a formal solution to (\ref{e3}) via
exponentiation of the ``tangent vector'' identity 
$\frac{d}{dt}\equiv{\cal L}_{h}/i\hbar$. Thus,
\begin{equation} 
\label{f1}
g_{t} = \exp\left\{
-i(t-t_{0}){\cal L}_{h}/\hbar\right\}\circ g_{t_{0}},
\end{equation}
where $L_{\Delta t}\equiv e^{-i(t-t_{0}){\cal L}_{h}/\hbar}$ 
is the  Liouville propagator. Now,
${\cal L}_{h}\circ(f+g) ={\cal L}_{h}\circ f + {\cal L}_{h}
\circ g$, so $L_{\Delta t}$ is a linear operator on
the vector space of Weinberg observables. However, because
${\cal L}_{h}$ depends, via $h$, upon $\psi$, the object
$L_{\Delta t}$ is usually a nonlinear operator when acting on
wavefunctions. Therefore, one must be exceedingly careful to
distinguish the trivial pseudo--superposition
$$(f+g)_{t}(\psi,\psi^{*}) = 
f_{t}(\psi,\psi^{*})+g_{t}(\psi,\psi^{*}),$$ 
which is always valid, from the special {\em trajectorial\/} 
superposition property  
$$(\psi+\phi)(t) = \psi(t)+\phi(t).$$ 
This is valid when $h(\psi,\psi^{*})$ is a linear functional
in both slots\cite{lin}, but fails in general (one sees this
easily from (\ref{e22}), if $\hat{H}$ depends upon $\psi$
then we cannot add operators for different states).

\subsection{The classical propagator}
Using the identity $[g_{0},h_{0}]_{\rm W}=i\hbar
n\{G,H\}_{\rm PB}$, valid for functionals of type (\ref{e1}),
and the fact that $dn/dt=0$, we recover the well--known
classical result:
$$G_{t} = 
G_{t_{0}} + 
\{G_{t_{0}},H_{t_{0}} \}_{\rm PB}
(t-t_{0})
+ \frac{1}{2!}
\{\{G_{t_{0}},H_{t_{0}}\}_{\rm PB},H_{t_{0}}\}_{\rm PB}
(t-t_{0})^{2}+ \ldots.$$
Similarly, one can use (\ref{f1}) to expand a formal solution
for the classical Schr\"{o}dinger equation. Here there is no
need given the exact solution (\ref{e33}).

\subsection{The quantum propagator}
For quantum functionals, as defined by (\ref{e4}), we invoke
the identity $[g_{1},h_{1}]_{\rm W}= \langle\psi|
[\hat{G},\hat{H}] |\psi\rangle$, and (\ref{f1}) becomes: 
$$\langle \hat{G}\rangle_{t} =
\langle\hat{G}\rangle_{t_{0}} + 
\langle[\hat{G},\hat{H}]\rangle_{t_{0}}
(t-t_{0})/i\hbar
+\frac{1}{2!} 
 \langle[[\hat{G},\hat{H}],\hat{H}]\rangle_{t_{0}}
(t-t_{0})^{2}/(i\hbar)^{2} + \ldots.$$
Similarly, using $[\psi,h_{1}]_{\rm W}=\hat{H}|\psi\rangle$,
one gets $|\psi_{t}\rangle =e^{-i(t-t_{0})\hat{H}/\hbar}
|\psi_{t_{0}}\rangle$. In this special case the propagator
does not depend upon $|\psi_{t_{0}}\rangle$.

This property encodes the superposition principle. All complexity 
lies in the propagator, which happens to be independent of the 
initial condition for linear theory. More generally this is not 
the case. Treating function--valued curves $\psi(t)$ as 
``trajectories'', the overlap:
\begin{equation}
\label{met}
{\cal D}(\psi,\psi') = 1 - |\langle\psi|\psi'\rangle|^{2}
\;\;\mbox{where}\;\;
{\cal D}\in[0,1],
\end{equation}
need not be constant in time. Divergence, and the possibility 
of strong divergence (i.e. ``exponential'', in some sense), is 
thus permitted in the nonlinear sector. To formalize this 
notion one can look to extend the KS--entropy, or the 
classical Lyapunov exponent to Weinberg's theory\cite{rei} via 
use of the metric (\ref{met}) (see \cite{jp}, for its properties).

\subsection{Dynamical chaos in the interpolative regime?}
In the interpolative case, an explicit computation of the
iterated bracket (\ref{e16}) is prohibitive. Nevertheless, 
the {\em existence\/} of a formal solution permits direct
study of the formal computability properties of both the
classical and quantal dynamics. Ford et al.'s algorithmic
information theory approach\cite{for} to the study of
``quantum chaos'' might extend in this direction.

On the numerical front, one needs to ascertain when, and how,
exactly, quantum suppression of chaos is {\em switched off\/}.
Certainly, it must happen at some $\lambda\in[0,1]$. Since 
(\ref{e16}) has a Poisson bracket contribution for every
$\lambda \ne 1$, this is the candidate chaos
factory\cite{fei}.

\section{Interpolative eigenstates}
\subsection{The fundamental variational principle}
Weinberg has generalized the eigenstates of linear quantum
theory as stationary points of the normalized observables 
via the simple variational principle\cite{sw5},
\begin{equation}
\label{varp}
\delta \left(\frac{h(\psi,\psi^{*})}{n(\psi,\psi^{*})}
\right) = 0,
\end{equation}
which is equivalent to\cite{sw1}:
\begin{eqnarray}
\label{e36}
\delta_{\psi^{*}}\left(\frac{h}{n}\right) 
& = & 
\frac{1}{n}\delta_{\psi^{*}} h - 
\frac{h}{n^{2}}\delta_{\psi^{*}}n = 0,\\
\label{e37}
\delta_{\psi}\;\,\left(\frac{h}{n}\right) 
& = & 
\frac{1}{n}\delta_{\psi}\;\, h - 
\frac{h}{n^{2}}\delta_{\psi}\;\, n = 0.
\end{eqnarray}
In the linear case this reduces to the familiar result
$\hat{H}|\psi\rangle = E|\psi\rangle$. Given the form       
of (\ref{e21}), we expect a similar result for the
special interpolative observables (\ref{e10}).

\subsection{Some preliminary observations}
Suppose, first of all, that $\psi$ is a stationary point
of the Weinberg observable $h(\psi,\psi^{*})$. Then, if
$a(\psi,\psi^{*})$ is {\em any other\/} Weinberg observable,
we can use the definitions (\ref{e2}), (\ref{e36}) and
(\ref{e37}) to compute 
\begin{eqnarray}
[a,h]_{\rm W}& = &
\delta_{\psi}a\delta_{\psi^{*}}h -
\delta_{\psi}h\delta_{\psi^{*}}a \nonumber \\
& = & \frac{h}{n}\left(
\delta_{\psi}a\delta_{\psi^{*}}n -
\delta_{\psi}n\delta_{\psi^{*}}a\right) \nonumber\\
& = & \frac{h}{n}[a,n]_{\rm W}= 0.
\label{e38}
\end{eqnarray}
This property generalizes the obvious fact that an eigenstate
$\psi$ of the linear operator $\hat{H}$, must return
$\langle\psi| [\hat{A},\hat{H}] |\psi\rangle=0$, for all
$\hat{A}$. 

As an immediate consequence of (\ref{e38}) we deduce, via  
the expressions (\ref{e18}) and (\ref{e19}), that 
\begin{equation}
\label{e39} 
\langle\hat{H}^{\lambda}_{q}\rangle = 0,\mbox{    and    }
\langle\hat{H}^{\lambda}_{p}\rangle = 0,
\end{equation}
of necessity. 

\subsection{The interpolative eigenvalue equation}
To construct the stationarity conditions for (\ref{e21}), we
substitute $\delta_{\psi^{*}}h= \hat{H}_{\rm eff}^{\lambda}
|\psi\rangle$ into (\ref{e36}), identify $\delta_{\psi^{*}}
n=|\psi\rangle$, and obtain the eigenvalue equation 
\begin{equation} 
\label{e40}
\hat{H}_{\rm eff}^{\lambda}|\psi\rangle
= \langle\hat{H}_{\rm eff}^{\lambda}\rangle|\psi\rangle.
\end{equation}
Combining (\ref{e39}) with (\ref{e21}) we see that
\begin{equation}
\label{e41}
(1-\lambda)\left\{\langle H^{\lambda}_{q}\rangle
(\hat{q}-\langle\hat{q}\rangle) +
\langle H^{\lambda}_{p}\rangle 
(\hat{p}-\langle\hat{p}\rangle)\right\}\equiv 0,
\end{equation}
which reduces (\ref{e40}) to
\begin{equation}
\label{e42}
\hat{H}^{\lambda}|\psi\rangle = E^{\lambda}|\psi\rangle,
\end{equation}
with the deformed eigenvalue,
$$E^{\lambda}\equiv
\langle\hat{H}_{\rm eff}^{\lambda}\rangle
= \langle\hat{H}^{\lambda}\rangle.$$
So (\ref{e40}) implies (\ref{e42}). Passing in the other
direction, we assume that $\lambda\ne0$, and notice that:  
$$\lambda \langle
\hat{H}^{\lambda}_{q}\rangle = 
\langle[\hat{H}^{\lambda},\hat{p}]\rangle/i\hbar,\; \mbox{   
and    } \lambda \langle \hat{H}^{\lambda}_{p}\rangle = 
\langle[\hat{q},\hat{H}^{\lambda}]\rangle/i\hbar,$$
whence (\ref{e42}) implies (\ref{e39}),
(\ref{e41}), and thus (\ref{e40}). 

To treat the exceptional point $\lambda=0$, we invoke
(\ref{e39}) alone, and deduce that the classical stationary
states of the deformed dynamical system comprise all $\psi$
such that $\langle\hat{q}\rangle$ and $\langle\hat{p}\rangle$
lie at a fixed point of the classical Hamiltonian flow (as 
one might have guessed). Clearly, such states have infinite
degeneracy, with a deformed eigenvalue that is precisely 
the classical energy at the fixed point. 

\subsection{The general solution via linear quantum theory}  
Equation (\ref{e42}) is simpler than
(\ref{e40}), but there remains a bothersome difficulty in that
\begin{equation}
\label{e43}
\hat{H}^{\lambda}=
\hat{H}(\lambda\hat{q}+(1-\lambda)\langle\hat{q}\rangle,
        \lambda\hat{p}+(1-\lambda)\langle\hat{p}\rangle).
\end{equation}
Although the expectation values are stationary, we have
to solve (\ref{e42}) self--consistently.

To fix this trouble, we bootstrap from solutions of the
simpler, {\em linear\/}, eigenvalue problem,
\begin{equation}
\label{e44}
\hat{H}(\lambda\hat{q},\lambda\hat{p})|\psi\rangle
 = E^{\lambda}|\psi\rangle.
\end{equation}
Defining the new operators: $\hat{q}'\equiv\lambda\hat{q}$, 
and $\hat{p}'\equiv\lambda\hat{p}$, we observe that $[\hat{q}',
\hat{p}']=i\hbar'$ with $\hbar'=\lambda^{2}\hbar$. Equation
(\ref{e44}) is, therefore, just the standard eigenvalue 
problem with a rescaled value of $\hbar$.

Given a parametric family of $\hbar$--dependent eigenstates
$\psi(q;\hbar)$, eigenvalues $E(\hbar)$, and eigenstate
expectations, $Q(\hbar)$, and $P(\hbar)$, for the ordinary
Schr\"{o}dinger problem, we identify:   
\begin{eqnarray*}
\label{e45}
\hbar'   & \mapsto & \hbar' = \lambda^{2}\hbar\\
\label{e46}
\hat{q}' & \mapsto & q'     = \lambda q\\
\label{e47}
\hat{p}' & \mapsto & -i\hbar'\partial_{q'} =
-i(\lambda^{2}\hbar)\partial_{(\lambda q)} =
\lambda(-i\hbar\partial_{q}).
\end{eqnarray*}
Thus the solution to (\ref{e44}) is obtained by applying the
rescalings $q\mapsto \lambda q$ and
$\hbar\mapsto \lambda^{2}\hbar$ to the known solutions for the
$\lambda=1$ problem. Imposing the constraint,
$\int_{-\infty}^{\infty} \psi(\lambda q)\psi^{*}(\lambda
q)\,dq=1$, now fixes the renormalized quantities:  
\begin{eqnarray}
\label{eigen}
\langle q|\psi_{\lambda}\rangle & = & \lambda^{1/2}
\psi(\lambda q;\lambda^{2}\hbar),\\
\label{energy} 
E^{\lambda} & = & E(\lambda^{2}\hbar),\\
\label{pos}
Q^{\lambda} & = &
\langle\psi_{\lambda}|\hat{q}|\psi_{\lambda}\rangle =
Q(\lambda^{2}\hbar)/\lambda,\\
\label{mom}
P^{\lambda} & = & 
\langle\psi_{\lambda}|\hat{p}|\psi_{\lambda}\rangle =
P(\lambda^{2}\hbar)/\lambda.
\end{eqnarray}
Using these expressions we can construct a solution to
the general problem (\ref{e42}). 

First we form, after (\ref{e25}), and using (\ref{pos})
and (\ref{mom}), the Weyl operator,
\begin{equation}
\label{e49}   
\hat{V}\equiv 
\hat{U}[(1-\lambda)Q^{\lambda},(1-\lambda)P^{\lambda}].
\end{equation}
Applying this to both sides of (\ref{e44}) gives, 
$$\hat{V}^{\dagger}
\hat{H}(\lambda\hat{q},\lambda\hat{p})\hat{V}
\hat{V}^{\dagger}|\psi_{\lambda}\rangle
= E^{\lambda}
\hat{V}^{\dagger}|\psi_{\lambda}\rangle.$$
Thus we can identify,
\begin{equation}
\label{e50}
|\psi'_{\lambda}\rangle
= \hat{V}^{\dagger}|\psi_{\lambda}\rangle
\end{equation}
as an eigenstate of the new operator, $\hat{V}^{\dagger}
\hat{H}(\lambda\hat{q},\lambda\hat{p})\hat{V}$ with the
eigenvalue $E^{\lambda}$ unchanged.

Using (\ref{e26}), (\ref{e27}) and (\ref{e50}) we compute:
\begin{eqnarray}
\label{e51}
\langle\psi'_{\lambda}|\hat{q}|\psi'_{\lambda}\rangle/n
& = & \langle\psi_{\lambda}|\hat{q}-
(1-\lambda)Q^{\lambda}
|\psi_{\lambda}\rangle/n = \lambda Q^{\lambda},\\
\label{e52}
\langle\psi'_{\lambda}|\hat{p}|\psi'_{\lambda}\rangle/n
& = & \langle\psi_{\lambda}|
\hat{p}-(1-\lambda)P^{\lambda}
|\psi_{\lambda}\rangle/n = \lambda P^{\lambda}.
\end{eqnarray}
Similarly,
\begin{eqnarray*}
\lefteqn{\hat{V}^{\dagger}
\hat{H}(\lambda\hat{q},\lambda\hat{p})\hat{V}
 = }\\ & &
\hat{H}(\lambda[\hat{q}+(1-\lambda)Q^{\lambda}],
\lambda[\hat{p}+(1-\lambda)P^{\lambda}]).
\end{eqnarray*}
Combining these relations, and comparing to (\ref{e43}), we
verify that solves (\ref{e42}) self--consistently. To pass in
the other direction, we start with a solution to (\ref{e42}),
pick $\hat{V}$ as the inverse of (\ref{e49}), with
$Q^{\lambda}$ and $P^{\lambda}$ determined from (\ref{e51})
and (\ref{e52}), and obtain, via (\ref{e50}), a solution of
(\ref{e44}). 

Making use of (\ref{eigen}), (\ref{pos}), (\ref{mom}) and
the disentanglement relation (\ref{dis}), 
\begin{eqnarray}
\lefteqn{\psi_{\lambda}(q) = \lambda^{1/2} 
e^{-i(1-\lambda)
P(\lambda^{2}\hbar)(\lambda q)/
(\lambda^{2}\hbar)} }\nonumber \\ & &
\hspace{2cm}\times
\psi(\lambda q + (1-\lambda)Q(\lambda^{2}\hbar);
\lambda^{2}\hbar),
\label{final}
\end{eqnarray}
where  all indicated functions are obtained as solutions
to the standard Schr\"{o}dinger problem ($\lambda=1$).

Recall the harmonic oscillator wavefunctions\cite{mor},
\begin{equation}
\label{hstates}
\psi_{n}(q;\hbar) = (2^{n}n!)^{-1/2}(\beta/\pi)^{1/4}
e^{-\beta q^{2}/2} H_{n}(q\beta^{1/2}),
\end{equation}
where $H_{n}(z)=(-1)^{n}e^{z^{2}}(d^{n}/dz^{n})e^{-z^{2}}$, 
with $\beta(\hbar) = m\omega/\hbar$. Since the position
and momentum expectations of these vanish, it is easy to
verify that (\ref{hstates}) are invariant under the 
transformation (\ref{final}). 

Although (\ref{final}) looks singular at $\lambda=0$, this
need not always be the case. As a matter of curiosity, we
wonder which class of Hamiltonians have eigenstates that 
are fixed points of this abstract mapping.

\subsection{Degeneracies and the failure of orthogonality}
Some minor trouble arises if (\ref{e44}) is degenerate. Then
(\ref{e49}) must be applied, in turn, to each member of the
invariant subspace associated with $E^{\lambda}$, so as to
generate a corresponding interpolative eigensubspace. Thus
one can think of the solutions to (\ref{e42}) as being
constructed by applying the {\em nonlinear\/} mapping
(\ref{e49}) to the entire Hilbert space. Evidently, the usual
linear eigenvector orthogonality relations are preserved, if,
and only if, all eigenvectors of (\ref{e44}) happen to share
identical coordinate expectations. Although the form of
$E^{\lambda}$ suggests, on first sight, that we are merely
taking $\hbar\rightarrow 0$ via a circuitous route, the
failure of orthogonality shows that the two approaches are, 
in fact, fundamentally different. One distinguishes this     
limit from the standard classical limit via the modification
to {\em eigenfunctions\/} (examine (\ref{final})). 

Another clear distinguishing feature is that we cannot
superpose the nonlinear eigensolutions 
$$|\psi'_{\lambda}(t)\rangle = 
e^{-i(t-t_{0})E^{\lambda}/\hbar}
|\psi'_{\lambda}(t_{0})\rangle,$$ 
to get a solution of (\ref{e22}).

\subsection{A connection between quantum eigenstates and \\
{\em fixed points\/} of the classical Hamiltonian flow?}
Given that $\lambda=0$ eigenstates lie at fixed points of  
the classical Hamiltonian flow, we conjecture that:
\begin{eqnarray}  
\label{lim1} \lim_{\lambda\rightarrow 0}
E(\lambda^{2}\hbar)  & = & E^{0}_{\rm f.p.},
\\ \label{lim2}
\lim_{\lambda\rightarrow 0} Q(\lambda^{2}\hbar)  & = &
Q^{0}_{\rm f.p.},\\ 
\label{lim3} \lim_{\lambda\rightarrow 0}
P(\lambda^{2}\hbar)  & = & P^{0}_{\rm f.p.}.
\end{eqnarray}
Two problems confound a proof. Firstly, continuity of the  
{\em defining\/} variational problem, (\ref{varp}), is
essential, but the infinite degeneracy of solutions at 
$\lambda = 0$ contradicts this. Secondly, at this same point
the {\em auxilliary\/} problem, (\ref{e44}), is  obviously
singular. So the known $\lambda=0$ behaviour need not always
connect with the above limits. 

For example, parity arguments applied to the quartic double
well potential, $V(q) = (q^{2}-1)(q^{2}+1)$, show that
(\ref{lim2}) fails. Eigenstates have vanishing expectation so
the two stable fixed points are missed out. Either conditions
of broken symmetry must obtain, or the correct statement is
more subtle.

For exact single fixed point problems, the limit (\ref{lim1})
is easily verified\cite{ver}. The harmonic oscillator obeys
it,  $$E^{\lambda}=\lambda^{2}\hbar\omega(n+1/2)
\rightarrow E^{0}=0,$$ 
as does the hydrogen atom,
$$E^{\lambda}_{n}= -\frac{Z^{2}e^{4}m_{e}}
{2n^{2}\lambda^{4}\hbar^{2}}\rightarrow E^{0}=-\infty,$$
(if we treat the origin as a fixed point). A soluble example
with two fixed points is Calogero's problem\cite{cal},
$$\left\{-\alpha\frac{\partial^{2}}{\partial q^{2}} 
+\beta q^{2} + \gamma q^{-2}\right\} 
\psi(q) = E\psi(q),$$ 
with the eigenfunctions\cite{abs},
$$\psi(q)=(\kappa q)^{a+1/2}e^{-\kappa^{2}q^{2}/2}
L^{a}_{n}(\kappa^{2}q^{2}),\;\;n=0,1,2,\ldots.$$ 
where $\kappa=(\beta/\alpha)^{1/4}$, $a=1/2(1+4\gamma/ 
\alpha)^{1/2}$, and $4\gamma/\alpha>-1$. The classical
fixed points lie at $q=\pm(\gamma/\beta)^{1/4}$, with
energy $2(\gamma/\beta)^{1/2}$. Taking Calogero's 
eigenvalue formula
$$E_{n} =(\alpha\beta)^{1/2}(2+2a+4n),$$ 
we let $\alpha\rightarrow 0$ and verify (\ref{lim1}).

Thus the energy result seems quite general. Indeed one can  
take the EBK semiclassical quantization rule\cite{ebk},
$\oint_{\Gamma} p\,dq = 2\pi\hbar(n+\alpha/4)$ and deduce
that, as $\hbar\rightarrow 0$, the symplectic area enclosed 
by the classical periodic orbits $\Gamma_{n}(\hbar)$ must
vanish. Now we assume that a {\em continuously\/} parametrized 
family of periodic orbits with this property must converge
upon {\em some\/} classical fixed point. Then EBK connects a
quantized energy level with the action parameter labelling the
``disappearing torus''. It appears that integrable Hamiltonians
must respect (\ref{lim1}).

\section{Uncertainty products and dispersion} 
\subsection{Generalized dispersion}
To develop a generalized uncertainty relation we recall
the usual definition, $\Delta^{2}_{a}\equiv\langle\psi|
(\hat{A}-\langle\hat{A}\rangle)^{2}|\psi\rangle$, where 
$\hat{A}$ is a {\em linear\/} operator. Then for
$a\equiv\langle\psi|\hat{A}|\psi\rangle$, we observe that
\begin{equation}
\label{a1}
\Delta^{2}_{a} = a\star a - a^{2}/n,
\end{equation}
where $a\star a\equiv\delta_{\psi}a\delta_{\psi^{*}}a$. If
we assume that $a$ commutes with all its $\star$--product
powers, then, Weinberg argues\cite{sw6}, the usual
probability interpretation is retained. Thus $a\star a$ is 
the average of the square, $a^{2}$ the average squared, and
(\ref{a1}) is a generalized {\em dispersion observable\/}. 

\subsection{A generalized uncertainty principle?}
Given a second observable $b$, whose $\star$--powers again
commute, we treat $\delta_{\psi^{*}}a$ and $\delta_{\psi^{*}}b$
as kets, their adjoints as bras, and set  
$$|\alpha\rangle = \delta_{\psi^{*}}a -
a/n\delta_{\psi^{*}}n, 
\mbox{    and    }
  |\beta\rangle  = \delta_{\psi^{*}}b - 
b/n\delta_{\psi^{*}}n.$$
Substituting these into the Schwartz inequality\cite{sch},
$\langle\alpha|\alpha\rangle
  \langle\beta|\beta\rangle \ge
 |\langle\alpha|\beta\rangle|^{2}$,
we collect $\star$--products to obtain the inequality
\begin{equation}
\label{a2}
(a\star a\!-\! a^{2}/n)
(b\star b\!-\! b^{2}/n)
\!\ge\! \left|(a\star b\! -\! ab/n)\right|^{2}.
\end{equation}
Working on the right hand side, we have 
$$a\star b - ab/n = \frac{1}{2}[a,b]_{\rm W} +
\frac{1}{2}[a,b]^{+}_{\rm W} - ab/n,$$
with $[a,b]^{+}_{\rm W}\equiv a\star b + b\star a$. Taking
the square norm, we observe that $1/2[a,b]_{\rm W}$ is pure
imaginary, while $1/2[a,b]^{+}_{\rm W} - ab/n$, is pure real.
Given that the real term vanishes on the minimum uncertainty
states, (\ref{a2}) permits the simpler, weakened, form
\begin{equation}
\label{a3}
\Delta^{2}_{a}\Delta^{2}_{b}\ge \frac{1}{4}
\left|[a,b]_{\rm W}\right|^{2}.
\end{equation}
Although this inequality bears a striking resemblance to the
standard Heisenberg--Robertson relation\cite{rob}, it is only
properly motivated if $a$ and $b$ are observables whose
$\star$--powers commute. Caution is advisable since the right
and left members of (\ref{a2}) need not be invariant under
general nonlinear canonical transformations. 

Although (\ref{a2}) has the formal properties of dispersion,
its physical interpretation is unclear. If dispersion depends
upon the coordinate system, we can make little of it, except
perhaps to distinguish the value zero as being special.

\subsection{A simple example: coordinate functionals}
For a simple example, we take the deformed coordinate
functionals $q_{\lambda}$ and $p_{\lambda}$. Since these
commute with their $\star$--powers, we have
$$\Delta^{2}_{q_{\lambda}}\Delta^{2}_{p_{\lambda}}
\ge \frac{1}{4}
\left|[q_{\lambda},p_{\lambda}]_{\rm W}\right|^{2} =
\frac{\hbar^{2}}{4}.$$ 
Thus deformation preserves the generalized uncertainty principle
(\ref{a3}), and coordinate dispersions are seen to obey the
usual interpretative rules.

\subsection{Wider validity?: classical observables}
Interestingly, the general stationarity conditions (\ref{e36})
and (\ref{e37}) imply, via (\ref{a1}), that dispersion must
vanish for generalized stationary states. This is the most
cogent physical reason for believing that (\ref{a2}) may be 
of general significance. 

For example, using (\ref{e12}) we compute,
$$\Delta_{h_{0}}^{2} =
(\partial_{q} H)^{2}\Delta_{q}^{2}
+2(\partial_{q} H)(\partial_{p} H)\Delta_{qp}^{2}
+(\partial_{p} H)^{2}\Delta_{p}^{2},$$
where,
$$\Delta_{qp}^{2}\equiv
\frac{1}{2}\langle\psi|
(\hat{p}-\langle\hat{p}\rangle)
(\hat{q}-\langle\hat{q}\rangle) +
(\hat{q}-\langle\hat{q}\rangle)
(\hat{p}-\langle\hat{p}\rangle)
|\psi\rangle.$$
Thus classical dispersion is just a ``quantized'' version of
gaussian quadrature error analysis. Dispersion vanishes at
classical fixed points, as does the right hand member of
(\ref{a3}) for quantities in involution (i.e. with zero 
Poisson bracket).

More generally the interpolative dispersion does not seem
to have any ready interpretation. We therefore doubt that
the concept is useful, except as a means to study the
spreading of quantum states under evolution. 

\subsection{Interpolative dynamics of dispersion} 
Since the generalized dispersions formed via rule (\ref{a1})
are again homogeneous of degree one, we can use the evolution
equation (\ref{e3}). For instance, from (\ref{e8}),
(\ref{e9}), and the formula (\ref{e16}) we compute
$[\Delta^{2}_{q_{\lambda}}, h_{\lambda}]_{\rm W}$ and
$[\Delta^{2}_{p_{\lambda}}, h_{\lambda}]_{\rm W}$, to obtain:
\begin{eqnarray} \label{a4}
\frac{d\Delta^{2}_{q_{\lambda}}}{dt} 
& = & +\lambda n \left\{
\langle[\hat{q},\hat{H}^{\lambda}_{p}]^{+}\rangle
-2\langle\hat{q}\rangle
\langle\hat{H}^{\lambda}_{p}\rangle\right\},\\
\label{a5}
\frac{d\Delta^{2}_{p_{\lambda}}}{dt} 
& = & -\lambda n \left\{
\langle[\hat{p},\hat{H}^{\lambda}_{q}]^{+}\rangle
-2\langle\hat{p}\rangle
\langle\hat{H}^{\lambda}_{q}\rangle\right\}.
\end{eqnarray}
No matter what the chosen state $\psi$, or Hamiltonian $H$,
dispersion is smoothly switched off as $\lambda\rightarrow 0$.

\section{The interpolative free particle}
To illustrate the preceding formal material we consider
the interpolative free particle Hamiltonian:
\begin{equation}
\label{i1}
\hat{H}^{\lambda}_{\rm eff}\equiv 
\frac{\hat{p}_{\lambda}^{2}}{2m} 
+\frac{\langle\hat{p}\rangle}{m}
(\hat{p}-\langle\hat{p}\rangle).
\end{equation}
Using either the propagator formula (\ref{f1}), or the fact
that the momentum $P_{0}=\langle\hat{p}\rangle$ is a constant
of the motion (via equations (\ref{e18}) and (\ref{e19})), we
see that the free particle propagator is just
\begin{equation}
\label{i2}
\hat{U}_{\Delta t} = \exp\left\{
\frac{-i\Delta t}{\hbar}\left(
a\hat{p}^{2}+b\hat{p} + c\hat{1}\right)\right\},
\end{equation}
where, from (\ref{i1}), the {\em constants\/} 
$a$,$b$ and $c$ read: 
\begin{equation}
\label{i3}
a=\frac{\lambda^{2}}{2m},\;
b = \frac{(1-\lambda^{2})P_{0}}{m},\;\mbox{and}\;
c = \frac{(\lambda^{2}-1)P^{2}_{0}}{m}.
\end{equation}
The problem is now easily solved using the deformed free
particle Green's function,  
\begin{eqnarray}
\lefteqn{K_{\lambda}(q',q;\Delta t) 
\equiv }\nonumber \\
& & \label{i4}
\frac{1}{2\pi\hbar}
\int_{-\infty}^{\infty}
e^{-i\Delta t\left(
ap^{2}+bp + c\hat{1}\right)/\hbar}
e^{+i(q'-q)p/\hbar}\,dp,
\end{eqnarray}
such that,
\begin{equation}
\label{i5}
\psi(q',t_{0}+\Delta t)
= \int_{-\infty}^{\infty}
K_{\lambda}(q',q;\Delta t)\psi(q,t_{0})\,dq.
\end{equation}
Evaluating (\ref{i6}) we get,
\begin{equation}
\label{i6}
K_{\lambda}(q',q;\Delta t) 
=(\pi/i\gamma)^{-1/2}e^{-i\kappa}
e^{i\gamma\left[q - (q'-\delta)\right]^{2}},
\end{equation}
where,
\begin{equation}
\label{i7}
\gamma =1/4a\hbar\Delta t,\;
\delta = b\Delta t,\;\mbox{and}\;
\kappa = c\Delta t/\hbar.
\end{equation}
Choosing an initial gaussian at the origin,
\begin{equation}
\label{i8}
\psi(q,t_{0}) = 
(\pi/2\alpha)^{-1/4}
e^{-\alpha q^{2} +i\beta q},
\end{equation}
with appropriate width and momentum parameters,
\begin{equation}
\label{i9}
\alpha =\frac{1}{4\sigma_{q}^{2}},\;\mbox{and}\;
\beta = \frac{P_{0}}{\hbar},
\end{equation}
we substitute (\ref{i6}) and (\ref{i8}) into (\ref{i5}),
and compute the evolved gaussian state,
\begin{equation}
\label{i10}
\psi(q,t_{0}+\Delta t)
= (\pi/2\alpha)^{-1/4}[(\alpha-i\gamma)/i\gamma]^{-1/2}
e^{-i[\kappa + \gamma(q-\delta)^{2}]}
\exp\left\{-\frac{\gamma^{2}[q-(\delta+\beta/2\gamma)]^{2}}
{(\alpha-i\gamma)}\right\},
\end{equation}
where primes are now dropped. Next we form, 
\begin{equation}
\label{i11}
|\psi(q,t_{0}+\Delta t)|^{2} =
\left(\frac{\pi(\alpha^{2}+\gamma^{2})}
{2\alpha\gamma^{2}}
\right)^{-1/2}
\exp\left\{-\frac{2\alpha\gamma^{2}
[q-(\delta+\beta/2\gamma)]^{2}}
{(\alpha^{2}+\gamma^{2})}\right\},
\end{equation}
and use (\ref{i3}), (\ref{i7}) and (\ref{i9}), to pick out the
evolved packet centre and dispersion formul{\ae}:
\begin{eqnarray} 
\label{i12}
q_{0}(t_{0}+\Delta t) & = &
\frac{P_{0}\Delta t}{m},\\
\label{i13}
\sigma_{q}^{2}(t_{0}+\Delta t) & = &
\sigma_{q}^{2}(t_{0})
\left\{1+\frac{\lambda^{4}\hbar^{2}(\Delta t)^{2}}
{4m^{2}\sigma_{q}^{4}(t_{0})}\right\}.
\end{eqnarray}
We check that interpolative particles propagate at the desired
classical velocity $P_{0}/m$. Moreover, as with the energies
$E^{\lambda}$, the formula (\ref{i13}) is identical to the
standard linear one, except that $\hbar$ is replaced by
$\lambda^{2}\hbar$. Compare the $\lambda = 0$ behaviour with
standard quantum theory. For any mass $m$, there exists some
time interval $\Delta t_{c}$, such that a particle will
eventually disperse so as to fill the entire known universe.
Ordinarily, we dispense with this difficulty by stating that
the interval is far too long to matter, and that particles
are, in any case, localized by measurements long before the
situation gets out of hand. In contrast, the limit (\ref{i13})
offers greater descriptive (not prescriptive) power in that we
can hang the value $\lambda=0$ upon this circumstance.

\section{Prospects for empirical test}
\subsection{Where does linearity apply, for sure?}
There have been numerous stringent tests of quantum linearity
performed upon {\em microscopic\/} systems. Each of these has
yielded a null result\cite{exp}. Bollinger et al\cite{exp},
have bounded the Weinberg nonlinearity in Beryllium nuclei
spin--precession experiments at less than $4$ parts in
$10^{-27}$. Other indirect tests, such as the atomic version
of Young's double slit experiment\cite{you}, and inversion
tunnelling in small molecules like Ammonia provide strong
evidence against nonlinearity in atomic scale systems.

\subsection{How might nonlinearity emerge?}
The quantum dynamics of isolated systems observed in today's
laboratory must therefore be linear to a very high degree of
precision. If nonlinearity lies somewhere, then it seems that
one must look for its effects in a {\em new\/} place. Either
that, or one argues that this {\em exact\/} version of
Hamiltonian classical dynamics, formulated as a {\em wave\/}
theory for {\em any\/} value of $\hbar$, is just a bizarre
mathematical accident, put there expressly to tease us.

A clear question emerges. Is quantum theory {\em always\/}
linear with an {\em approximate\/} classical limit; or is 
there a more general nonlinear theory which is linear for
small systems and progressively nonlinear until we recover 
an {\em exact\/} classical limit? 

Two distinct physical interpretations appear possible. Either
the $\psi$--dependent operators express a statistical result
that should then be traced to environment--induced fluctations
(decoherence\cite{dec}); or, since (\ref{e3}) is deterministic,
the nonlinearity might reflect a purely causal coupling to 
the environment (a back--reaction or self--energy effect). In
either case, it seems {\em plausible\/} that nonlinearity
should become larger the less isolated, and more entangled,  
a quantum system becomes.

\subsection{In search of a mesoscopic ``elementary particle''}
Most elementary particles have internal structure. However, 
if empirical energy scale is decoupled from the internal
degrees of freedom, then we can exploit a structureless
one--particle approximation. 

In particle physics one reveals internal structure by building
a higher energy accelerator. To test any one--particle wave
equation one needs an inverted version of this program. The 
goal is to screen the known internal degrees of freedom and 
get the detector energies {\em low\/} enough (or sideband
them on a more accessible frequency).

To make a mesoscopic ``elementary particle'' we could take a
spherical macromolecule, or perhaps a microsphere\cite{msp}.
Then we charge it, or magnetize it, and find an ingenious way
to measure this {\em and\/} weigh it\cite{wei}. Then we give
the particle a moment of some kind, put it in a well and
couple it to coherent radiation  in an accessible range
(probably microwaves). Then it is feasible,  in principle, to
resolve the quantized energy levels. Nobody does this now
because it seems impossible to get the thermal background cool
enough, or the characteristic frequencies high enough, to be
able to resolve the levels of a particle in, say, the
microgram range. The lighter our particle the easier the
experiment, but the further we are likely to be from the
classical regime.

\subsection{A possible empirical signature}
Suppose we can do this at some mass (or size) scale. Given the
standard prediction for energy levels $E(\hbar)$, one needs to
use the spectroscopic data, along with the known particle mass
etc., to {\em measure\/} Planck's constant (assuming the
radiation law $\Delta E=\hbar\nu$). If this were to exhibit
a {\em monotonic decrease\/} as one passes to more classical
systems, then has evidence for a perturbative energy level
shift, like the $E(\lambda^{2}\hbar)$ effect.  Because the 
classical and quantal Weinberg energy functionals differ, 
one might expect something similar for any interpolative 
scheme. Our investigation is thus helpful, if only to show 
that any observed discrepancy of this kind deserves careful 
attention.

\section{Theoretical difficulties}
Given that experimental tests of the validity of exact
linear quantum theory in the classical domain are so
very difficult; we now highlight some of the severe
problems the nonlinear theory generates. It may be
that strong exclusions can be found via this route.

\subsection{The free nature of $\lambda$}
This is the most obvious problem. Without positive empirical
evidence one cannot fix $\lambda$. The only thing we learn is
what kind of effects one might need to look for. There does
not seem to be any way around this problem. Remember also
that (\ref{e10}) is just a postulate. Canonical quantization
is not the only route to generalization.

\subsection{Lack of manifest algebraic closure}
From (\ref{e15}), we see that the interpolative observables
(\ref{e10}), do not manifestly comprise a subalgebra, except 
at $\lambda=0,1$. This ugly mathematical feature strongly
suggests that the interpolation is unphysical. A subalgebra 
may show up using coordinate free methods (i.e. write
(\ref{e10}) in terms of $\star$--products and ordinary
products). However, this fact, and the general complexity
of the interpolative domain, leads us to conclude that
(\ref{e10}) has no {\em fundamental\/} physical content, 
other than as a guide to formulating empirical questions.

At a deeper level we obtain a sieve: ``What existence and 
uniqueness constraints apply to a one--parameter family
of Weinberg subalgebras which joins the classical and
quantum regimes?''. 

\subsection{Problems with measurement: a provisional\\
probabilistic interpretation}
Weinberg has emphasized\cite{sw7} that generalization of the
probability interpretation to nonlinear observables is 
defeated by non--associativity of the functional
$\star$--product. Nor can we use the Hilbert space inner
product, since this is not a canonical invariant in the
nonlinear sector of the theory.

How else might we get a probability interpretation? Since
problems arise due to nonlinearity, the natural place to
look for the ``right'' idea is in this sector. It is much
easier to specialize a working result; than to generalize
from a special one. 

Classical statistical physics employs densities $\rho(q,p)$
on phase space. Liouville's theorem preserves normalization and
Hamilton's equations determine evolution of the ensemble. One
can then discuss {\em classical\/} measurement as a stochastic
diffusive process superimposed upon the dynamics, and justify 
statistical mechanics via the ergodic hypothesis\cite{lif}.

Classical expectations are phase space averages
\begin{equation}
\label{cavg}
\bar{f} =\int \rho(q,p) f(q,p) \,dp dq,
\end{equation} 
where $\rho(q,p)$ is stationary. 

Since Weinberg's theory specializes the Hamiltonian formalism
(homogeneity is a {\em constraint\/} upon the hamiltonian) we
can try and carry this over directly. The key is to find an
invariant measure upon quantum states.

Canonical invariance of the symplectic form $dp\wedge dq$ (and
thus its exterior powers), implies Liouville's theorem\cite{arn}. 
In Weinberg's theory we identify the corresponding canonically
invariant symplectic form $\sum_{k=1}^{D} d\psi^{*}_{j}\wedge 
d\psi_{j}$. Taking exterior powers of this we get Liouville's
theorem, and an induced invariant measure on the projective 
Hilbert space of normalized states.

Exploiting canonical invariance of the norm $n$, we focus
on functionals that are homogeneous of degree $p$, and 
define the measure\cite{kj1}:  
\begin{equation} 
\label{avg}
\int F(\psi,\psi^{*})\, d\hat{\Omega}_{\tilde{\psi}}
\equiv \frac{\Gamma (D)}{\Gamma(D+p)}
\int F(\psi,\psi^{*}) e^{-n}\,
\prod_{j=1}^{D} \pi^{-1}
d{\rm Re}[\psi_{j}]d{\rm Im}[\psi_{j}],
\end{equation}
where $d\hat{\Omega}_{\tilde{\psi}}$ emphasizes the analogy 
with solid angle. To get a good $D\rightarrow\infty$ limit, 
we set $p=0$ on the right hand side (divide $F$ by $n^{p}$ 
when taking the average). 

The formula (\ref{avg}) is immediately recognized as the
standard functional measure of path integrals\cite{sch},  
or the theory of gaussian random fields\cite{gud}.

Now let $\rho(\psi,\psi^{*})$ be any {\em positive\/}
Weinberg observable satisfying,
$$\int \rho(\psi,\psi^{*})\, d\hat{\Omega} = 1.$$
The uniform density becomes $n(\psi,\psi^{*})$, the norm
functional. A nontrivial example is,
\begin{equation}
\label{del}
\rho^{\phi}_{N}(\psi,\psi^{*}) = n^{1-N}
\frac{\Gamma(D+N)}{\Gamma(N)\Gamma(D)}
|\langle\psi|\phi\rangle|^{2N},
\end{equation}
where the factor $n^{1-N}$ makes this homogeneous of degree
one, so that (\ref{e3}) applies. On averaging we set this
to $n^{-N}$. As $N\rightarrow\infty$, (\ref{del}) peaks
strongly about $\phi$. This density plays the role of 
a delta function on states.

To see this, we use the formula\cite{kj2},
\begin{eqnarray}
\int |\langle\phi|\psi\rangle|^{2}
f(|\langle\omega|\psi\rangle|^{2})
\,d\hat{\Omega}_{\tilde{\psi}} & = &
\frac{1}{D-1}(1-|\langle\phi|\omega\rangle|^{2})
\int f(|\langle\psi|\omega\rangle|^{2})
\,d\hat{\Omega}_{\tilde{\psi}} \nonumber \\
& & +
\frac{1}{D-1}(D|\langle\phi|\omega\rangle|^{2} -1)
\int |\langle\psi|\omega\rangle|^{2}
f(|\langle\psi|\omega\rangle|^{2})
\,d\hat{\Omega}_{\tilde{\psi}}.
\label{form}
\end{eqnarray}
Defining generalized quantum averages in the classical fashion,
\begin{equation}
\label{abar}
\bar{a}=\int \rho(\psi,\psi^{*}) a(\psi,\psi^{*})
\, d\hat{\Omega}_{\tilde{\psi}},
\end{equation}
we choose the bilinear functional,
\begin{equation}
\label{spec}
a_{1}(\psi,\psi^{*}) = \langle\psi|\hat{A}|\psi\rangle
= \sum_{j=1}^{D} a_{j}
|\langle\psi|\omega_{j}\rangle|^{2},
\end{equation}
where $A_{j}$ are the eigenvalues of $\hat{A}$, and
$|\omega_{j}\rangle$ its eigenvectors. Substituting
(\ref{spec}) and (\ref{del}) into (\ref{abar}), we
use (\ref{form}) and the definition (\ref{avg})
to verify that,
\begin{eqnarray}
\lefteqn{\int \rho^{\phi}_{N}(\psi,\psi^{*}) 
a_{1}(\psi,\psi^{*})\,d\hat{\Omega}_{\tilde{\psi}} = } \\
& & \sum_{j=1}^{D}\left\{ \frac{D}{(D-1)(D+N)}
+ \frac{N}{D+N}\left(1 - \frac{D}{N(D-1)}\right)
A_{j} |\langle\phi|\omega_{j}\rangle|^{2}\right\}.
\end{eqnarray}
Keeping $D$ fixed, and taking $N\rightarrow\infty$, we
recover the desired result
\begin{equation}
\label{qavg}
\bar{a}_{1} = \int \rho^{\phi}_{\infty}(\psi,\psi^{*}) 
a_{1}(\psi,\psi^{*})\,d\hat{\Omega}_{\tilde{\psi}}
= \langle\phi|\hat{A}|\phi\rangle.
\end{equation}
Thus quantal expectation values can be reinterpreted 
as phase space averages with respect to the delta
distribution $\rho^{\phi}_{\infty}$. 

If we let $D\rightarrow\infty$, in heuristic fashion, and 
choose classical Weinberg functionals, then (\ref{avg})
induces the standard Liouville measure over the phase
space of coordinate expectations, and we recover the 
classical result (\ref{cavg}).
 
More generally, one consider $\rho(\psi,\psi^{*})$ as 
defining the density matrix,
\begin{equation}
\label{denisty}
\hat{\rho} = \int \rho(\psi,\psi^{*})
|\psi\rangle\langle\psi|\,d\hat{\Omega}_{\tilde{\psi}}.
\end{equation}
Linearity of the trace operation and positivity of
the probability density ensures that, $\hat{\rho}> 0$ 
and ${\rm Tr}[\hat{\rho}] = 1$. This connection is
many--to--one, so that $\rho(\psi,\psi^{*})$ is a 
``hidden'', or ``indeterminable'', representation 
of $\hat{\rho}$. Nevertheless, 
$$\bar{a}_{1} = 
\int \rho(\psi,\psi^{*}) a_{1}(\psi,\psi^{*})
\,d\hat{\Omega}_{\tilde{\psi}}
={\rm Tr}[\hat{\rho}\hat{A}].$$
Pure states become delta function ensembles, whereas smeared
densities generate the mixed states. 

Rephrasing quantum averages in this fashion, we acquire a 
common statistical language for both classical and quantum
physics. Next we need to incorporate a generalized theory of
measurement. Thusfar we can only do this by appeal to the
known result. Nevertheless, our hope is that a suitably
generalized perspective might reveal (\ref{jump}) as a 
special case, whose inner product nature is accidental 
to the linear sector, but somehow {\em necessary\/}.

For a quantum system in state $\phi$ subjected to a 
complete measurement with the operator $\hat{A}$, we 
have the jump process $\phi\mapsto\omega_{j}$ 
occuring with conditional probability,
\begin{equation}
\label{jump}
p(\omega_{j}|\phi) = 
|\langle\phi|\omega_{j}\rangle|^{2}.
\end{equation}
Post measurement, we have a probability density peaked
as delta function spikes on each of the eigenvectors,
with weight given by rule (\ref{jump}):
\begin{equation}
\label{meas}
\rho(\psi,\psi^{*}) = \sum_{k=1}^{D}
|\langle\phi|\omega_{j}\rangle|^{2}
\rho^{\omega_{j}}_{\infty}(\psi,\psi^{*}).
\end{equation}
Substituting this into the rule (\ref{abar}), we verify
that $\bar{a}=\langle\phi|\hat{A}|\phi\rangle$. This is
the same as the result (\ref{qavg}), but the underlying 
distribution over states is {\em different\/}.

How can we understand this? Although (\ref{del}), with
$N\rightarrow\infty$, and (\ref{meas}) generate exactly
the same statistics, their {\em dynamical\/} properties 
under (\ref{e3}) are different. If $\hat{A}$ were the
Hamiltonian then (\ref{meas}) is a {\em stationary\/}
probability density. In contrast, the density (\ref{del}) 
must be time--dependent, unless $\phi$ happens to be 
an eigenstate of $\hat{A}$.

Thus the stationary states of the hamiltonian flow appear
rather special to the quantum case, they are associated
with stationary probability densities which are sums of
delta functions upon these. This suggests that we should
incorporate (\ref{jump}) as a dynamical result\cite{bel},
albeit via {\em stochastic dynamics\/}\cite{not}. 

Master equations, either classical, or quantal, are the
canonical examples of this paradigm\cite{fok}. They
encapsulate stochastic evolution of individual ensemble 
members via a deterministic equation of Fokker--Planck 
type. Adopting this formal route, we postulate the
generalized nonlinear quantum master equation\cite{mil}:
\begin{equation}
\label{master}
\frac{d\rho}{dt}
= \frac{1}{i\hbar} [\rho,h]_{\rm W}
+ \frac{\Gamma}{(i\hbar)^{2}}
  [[\rho,a]_{\rm W},a]_{\rm W},
\end{equation}
where $h$ is the ``free evolution'' and $a$ is the
``measurement functional'', while $\Gamma$ is a
phenomenological parameter (zero if measurement 
is switched off). 

Given (\ref{master}) we must solve for the stationary
probability density $\rho_{\infty}$ (defined as the
limit $t\rightarrow\infty$) generated from a chosen
initial condition $\rho_{0}$. Assuming existence of
$\rho_{\infty}$, the averaging rule (\ref{abar}) 
provides the statistical prediction.

From (\ref{master}) the stationarity condition reads
$\dot{\rho}=0$. To recast this we introduce Liouville 
operators: ${\cal L}_{h} \equiv [\bullet,h]_{\rm W}$, 
and ${\cal L}_{a} \equiv [\bullet,a]_{\rm W}$, to get: 
\begin{equation}
({\cal L}_{h} - {\cal L}_{a}\circ{\cal L}_{a})
\circ \rho = 0.
\end{equation}
This specifies the kernel of a {\em linear\/} operator, 
\begin{equation}
\label{mop}
{\cal L}_{\rm M}\equiv 
{\cal L}_{h} - {\cal L}_{a}\circ{\cal L}_{a},
\end{equation}
on the space of Weinberg functionals (the operator acts in
the adjoint representation of this Lie algebra). Thus we
identify ${\cal L}_{\rm M}$ as the formal ``measurement 
operator'' which is to describe an $a$--measurement
perfomed upon a system undergoing $h$--evolution.

Conveniently, linearity of (\ref{mop}) implies a spectral
theory. Intuitively, we expect the spectrum of (\ref{mop}) 
to determine the decay rate of $\rho_{0}$ to $\rho_{\infty}$,
and also the class of initial conditions $\rho_{0}$, upon
which a given measurement will be good (in the sense that 
we get to a stationary density, or arbitrarily close to it, 
in a finite interaction time). If $\rho^{k}_{\infty}$ is
a finite set $k\in[1,M]$ of stationary densities, then
so too is the linear combination,
\begin{equation}
\label{rhosup}
\rho_{\infty} = \sum_{k=1}^{M}w_{k}\rho^{k}_{\infty},
\end{equation}
provided only that the weights $w_{k}$ sum to unity. Thus
a measurement theory somewhat analagous to that of linear
quantum theory exists, even when orthogonality is relaxed
(recovery of this, on the space of density functionals,
would require (\ref{mop}) to be self--adjoint).

The generalization is certainly suggestive. Significantly, the
equation (\ref{master}) is not a hamiltonian flow, but it has
the desired physical property of being expressed purely via
canonically invariant Weinberg brackets.

Unsolved problems aside, a consistent, and inclusive,
statistical interpretation of nonlinear quantum theory  
is {\em conceivable\/} via appeal to stochastic dynamics. 

\subsection{Thermodynamic constraints}
Although $\hbar$ is fixed, harmonic oscillator energy levels
have the reduced spacing $\lambda^{2}\hbar\omega$ and the
$n$th stationary solution now reads:
\begin{equation}
\label{hosc}
|n_{t}\rangle=
e^{-i\lambda^{2}\omega(n+1/2)(t-t_{0})}
|n_{t_{0}}\rangle.
\end{equation}
However, from (\ref{e18}) and (\ref{e19}) one verifies that
a gaussian wave packet oscillates at the classical frequency
$\omega$, for all $\lambda$. Thus we encounter the bizzare
circumstance that a transition between two stationary states
suggests the photon frequency $\nu=\lambda^{2}\omega$, while
the classical radiation frequency remains $\omega$.

We could try and fix this by letting $\nu=\omega$ so that the
photon energies become $\lambda^{2}\hbar\nu$. However, that
leads to the horrible consequence that photons must either, be
confined to a single $\lambda$--sector, or, change their
frequency at each interaction with matter. Worse still, the
deformed Planck black body factor (excluding degeneracy),
\begin{equation}
\label{planck}
\frac{e^{-\lambda^{2}\hbar\nu/kT}}    
{1 -e^{-\lambda^{2}\hbar\nu/kT}},
\end{equation} 
detonates at $\lambda = 0$. Presto, an ultraviolet catastrophe!
The only way to salvage this disaster is to postulate that
photons are {\em always\/} $\lambda = 1$ particles. To defend
that, superposition of light waves is regularly observed at
the classical level, whereas that of matter waves is not.

Thus a deformed harmonic oscillator {\em must\/} have two
characteristic frequencies. This curious property offends
cherished physical intuition. However, as shown in great depth
by Weinberg\cite{sw8}, such behaviour is common. Moreover,
because the state preparations differ, it would be impossible
to observe both frequencies in a single experiment. Evidently,
there is no ambiguity or contradiction, a situation not unlike
wave--particle duality. How one could ever detect this is a
problem, unless perhaps thermodynamics can do it for us via
some modification of specific heats. Certainly, the deformed
black body rule would predict this; but given the historical
importance of that problem it is hard to believe that there
is any discrepancy lurking in the data. Currently it is
assumed that material and radiative oscillators must be
quantized in the same way. Certainly radiation must be
consistently quantized, because it mediates interaction
between material particles. That leaves us in some doubt
as to whether material oscillators {\em must\/} obey the
same rule of consistency. Clarification of this issue is
probably the most powerful constraint upon any modified
quantum theory. Nobody would reject thermodynamics.

\section{Conclusion}
In summary, we have embedded both Hamiltonian classical mechanics
and linear quantum theory as two disjoint dynamical sectors of
Weinberg's generalized nonlinear theory. To explore the idea
of a mesoscopic regime we then studied one technique for
interpolation. Although not fully constrained, our method is
simple, general, and has some desirable physical features. 
The result is an alternative classical limit whereby quantal
evolution is smoothly transformed into classical evolution as
we vary a single dimensionless control parameter $\lambda$.
Significantly, this works for any value of $\hbar$.

At the level of mathematical physics, we have a new tool for
comparing classical and quantal dynamics.  This can be put
to immediate use in studies of ``quantum chaos''. The ability
to turn dynamical chaos on and off via $\lambda$, whatever the
magnitude of $\hbar$, provides a new probe of the origin of
dynamical chaos suppression, and the potential for exposing
some interesting phenomena in the transition regime. We will
return to study this later, with a parting comment that the
interpretative problems have no bearing upon this pursuit.

Concerning the working hypothesis that nonlinearity emerges 
at the classical level, we stress that the evident success of 
linear quantum theory for microscopic systems is not in dispute.
Rather we imagine that a complex of atomic systems,  a whole
molecule, a block of solid, glass of beer, cat, flea on cat,
or ribbon of its DNA, has gotten complicated enough that the
dynamics for the $\psi$ of its centre of mass is described  
by a nonlinear theory. 

This attempt at a physical interpretation is imprecisely
formulated. The mathematics is unwieldy, and devoid of 
predictive power. Given its complexity, we do not believe 
that the interpolative technique has any fundamental 
physical content. Nevertheless, the one--particle 
assumption at least enables us to compute deformed
energies $E(\lambda^{2}\hbar)$, and show that the 
free particle has uniformly suppressed dispersion
as $\lambda\rightarrow 0$. Thus we settle upon the
view that the proper role of interpolative dynamical 
studies is to guide tests of the {\em universality\/} 
of linear quantum theory. 

Fundamental questions of this nature demand careful 
scrutiny. Indeed, the idea of emergent nonlinearity, 
bizarre as it may be, is consistent with both the 
observed linearity of isolated atomic scale systems 
and the fact that classical mechanics describes the
familiar world of our senses so well. In the current
climate one is led to reject a complete recovery of 
classical theory, because it {\em implies\/} that 
there is a nonlinear regime, and so linear quantum 
theory could not be considered universal. We suggest 
that if our prejudices demand that we invent reasons 
to ignore simple mathematical facts, then physics 
is in very serious trouble.

\section{Acknowledgments}
Portions of this work were carried out at the University of
Melbourne, Australia; University of Houston, Texas; University
of Texas at Austin; Institute of Advanced Study Princeton; and
my current address. I am grateful to: B.H.J. McKellar, A.G.
Klein, S. Adler, S. Weinberg, S.C. Moss, and M. Eisner for
their hospitality, and for useful discussions. Conversation or
correspondence with: A.J. Davies, S. Dyrting, O. Bonfim, N.E.
Frankel, Z. Ficek, G.J. Milburn, V. Kowalenko, H. Wiseman,  
R. Volkas and I.C. Percival sharpened the ideas, and provided
encouragement. Support from the University of Melbourne, A.G.
Klein, B.H.J. McKellar and N.E. Frankel, via a Visiting
Research Fellowship, and a Special Studies Travel Grant are
further acknowledged, along with an A.R.C. postdoctoral
fellowship.


\begin{references}
\bibitem{sup} If $\psi_{1}(t)$ and $\psi_{2}(t)$ are solutions
then so too is $\psi(t) = \alpha\psi_{1}(t)+\beta\psi_{2}(t)$,
even if these states are non--orthogonal. The normalisation is
time independent; absorb it into $\alpha$ and $\beta$. Then,
using results from Ref. \cite{lim}, one can choose states
such that $\langle\hat{A}\rangle_{\psi_{1}}(t)$ and
$\langle\hat{A} \rangle_{\psi_{2}}(t)$ follow the classical
trajectories. But now $$\langle\hat{A}\rangle_{\psi}(t) =
|\alpha|^{2} \langle\hat{A}\rangle_{\psi_{1}}(t) +  |\beta|^{2}
\langle\hat{A}\rangle_{\psi_{2}}(t) + 
2{\rm Re}[\alpha^{*}\beta\langle\psi_{1}(t)|\hat{A}
|\psi_{2}(t)\rangle],$$
which need not follow any classical trajectory.
\bibitem{cat} The paradox of Schr\"{o}dinger's cat is thus
avoided by the practical assertion that non--local, alive
and dead or correlated singlet felines are unpreparable.
\bibitem{fun} All computations follow from the bilinear
result $\delta_{\psi}\langle\psi|\hat{A}|\psi\rangle  
=\langle\psi|\hat{A}$, where $\hat{A}$ is any linear operator.
This device, from Ref.\cite{jon}, permits direct comparison
with standard theory. As per Ref.\cite{sw3} inner products
apply between quantities defined at the same $\psi$.
\bibitem{eva} This is why we adopt the interpretation noted
at Ref.\cite{lie}. The $Q(t)$ and $P(t)$ are not precisely 
measurable, but they can ``guide'' $\psi$ along a classical 
path.
\bibitem{lin} It is a folk prejudice of the quantum chaos
community that the linearity of quantum theory has nothing to
do with chaos suppression because the classical Liouville
equation has this trivial linearity property. Wider study of
nonlinear quantum theory, via numerical simulations, should
help decide the matter.
\bibitem{wei} For true elementary particles the requisite
parameters can be measured in different experiments, for a
composite particle it becomes very much harder.
\bibitem{sw8} Many characteristic effects of this kind are
developed in Ref. 11 \S 6. Although our interpolation does not
lie in the class considered by Weinberg, the stringent
exclusions made there appear to extend to this work also.
\end{references}
\end{document}